# Towards a Multi-Criteria Development Distribution Model: An Analysis of Existing Task Distribution Approaches

Ansgar Lamersdorf
*University of Kaiserslautern*
*a_lamers@informatik.uni-kl.de*

Jürgen Münch
*Fraunhofer IESE*
*Juergen.Muench@iese.fraunhofer.de*

Dieter Rombach
*University of Kaiserslautern and Fraunhofer IESE*
*Dieter.Rombach@iese.fraunhofer.de*

**Abstract**

*Distributing development tasks in the context of global software development bears both many risks and many opportunities. Nowadays, distributed development is often driven by only a few factors or even just a single factor such as workforce costs. Risks and other relevant factors such as workforce capabilities, the innovation potential of different regions, or cultural factors are often not recognized sufficiently. This could be improved by using empirically-based multi-criteria distribution models. Currently, there is a lack of such decision models for distributing software development work. This article focuses on mechanisms for such decision support. First, requirements for a distribution model are formulated based on needs identified from practice. Then, distribution models from different domains are surveyed, compared, and analyzed in terms of suitability. Finally, research questions and directions for future work are given.*

## 1. Introduction

Distributed or Global Software Development (GSD) has become more and more common practice for software development projects ([25], [13], [21]) due to its expected advantages. However, in many cases the crucial factor for the decision of starting GSD is the cost argument: Activities are outsourced to companies or regions (such as Eastern Europe or Asia) where wages are lower. Examples are described in [8] and [42].

This strategy bears a set of risks due to the specific problems of distributed development, such as decreased efficiency regarding communication, coordination, and control ([22], [7], [31], [24]), lack of trust ([42], [8]), or insufficient knowledge about remote sites ([8], [36]).

These problems are expected to have a negative impact on development productivity and the resulting quality of the software, making the overall development costs higher, even if the hourly costs are lower ([3]). Some voices even argue that these problems outweigh the benefits of GSD ([43]).

On the other hand, there are other goals in a software development project that might benefit from GSD, such as higher product quality through better knowledge at other sites or shorter development time through "follow-the-sun" development ([7]).

Therefore, a tradeoff analysis between different (potentially conflicting) project goals for GSD is necessary. This obviously depends on the assignment of the different development tasks to the available sites. For instance, assigning two tasks that depend very much on each other (e.g., the design of two highly coupled subsystems) to two different sites might lead to many communication problems. These could be reduced by assigning these tasks to one site while assigning other tasks with lower coupling to different sites (e.g., the design of two subsystems is done at one site and testing at another one). Depending on the goals and context, there might be many other factors, such as available expertise and capacity, which might influence the decision about assigning tasks to sites.

This decision process could be improved with the use of a decision support system. Such a system must be built on a distribution model that models the important characteristics of a project and of the development sites in order to evaluate assignments. Nowadays, there is a lack of systematic decision support for distributing tasks in GSD project planning. However, some initial distribution models exist in GSD and other domains.

In this article, selected distribution models are identified, analyzed, and compared with respect to their applicability for a decision support system for GSD. The remainder of this article is structured as follows: Section 2 introduces the concept of decision support systems and distribution models and reports related experience in GSD. The requirements for a distribution model are then derived in section 3. Section 4 presents existing task distribution approaches from several domains. In section 5, these approaches are compared with respect to the stated requirements. Finally, section 6 summarizes the findings and gives an outlook on future work. The Appendix provides a rough sketch of how a distribution model from distributed systems could be applied to GSD.

## 2. Background

In the following, basic terminology and concepts regarding decision support systems as well as experience from practice with respect to challenges of distributing work in GSD are described.

### 2.1. Decision Support System and Distribution Model

Scott-Morton [40] defines a *Decision Support System* (DSS) as an "interactive computer-based system, which helps decision makers utilize data and models to solve unstructured problems". The term "unstructured problems" refers to complex, unclear problems for which no standard solutions exist [45]. The assignment of activities to distributed development sites can be seen as a complex problem that could be supported by a DSS.
A DSS consists of three main subsystems [45]: Data management, dialog management, and model management. In the remainder of this paper, we will focus on creating models for a DSS that reflect the distribution of activities across sites. We therefore call them *Distribution Models*. A distribution model contains the following items:
*Goal*: Criteria that define states to be achieved (e.g., with respect to cost or development time). Usually there are different, sometimes conflicting goals.
*Tasks*: A set of work units (e.g., process steps, roles) that can be assigned to different sites. So-called characteristics can be attributed to the work units.
*Resources*: The set of available resources at all sites. A site is a locally and organizationally coherent group of resources that is separated from other sites.
*Mapping*: The assignment of tasks to sites. In a complete mapping, every task is assigned to one resource at one site.

The distribution model can now be seen as a function:

$$f: (Tasks \times Resources \times Mapping) \rightarrow Goal^n$$

The objective of the DSS can be defined as maximizing the goal vector.

Not much has been reported yet about modeling distribution for finding optimal assignments in GSD. Some approaches exist, though, such as in [44]. These approaches typically lead to quite general statements about distribution strategies. For supporting concrete distribution decisions, these approaches provide only limited value. However, many distribution models exist in other domains, as we will show in section 4.

### 2.2. Experiences from Practice: Problems and Challenges in GSD

The following section reports on the state of the practice in GSD based on a literature review.

When GSD is initiated by outsourcing software development, it is usually driven by different goals. In IT outsourcing, the top four goals according to Gareiss [18] are cost savings, additional expertise, staffing issues, and increased flexibility. The goal of cost savings is by far the most important factor driving more than 60 percent of the outsourcing relationships. However, in recent years, other goals such as development quality have gained more importance, which can be seen by the fact that service providers for GSD now advertise their quality improvement more than their low costs ([26]).

But cost savings cannot be achieved easily, even if the wages are lower: According to [18], nearly 50 percent of the companies doing outsourcing needed more management time than expected; more than 40 percent experienced insufficient performance; and nearly 40 percent reported additional unexpected costs.

The reasons for GSD projects and software development outsourcing being unsuccessful resemble each other very much. In general, they can be classified into two groups: problems *between different sites* and problems *at sites*.

The problems between different sites are inspired by the distributed nature of GSD. This has various reasons such as:

- *Communication barriers*, complicating communication between sites and therefore decrease the overall efficiency ([22], [24], [31], [12]),
- *Time shift* between time zones, reducing the overlap time and the possibilities of working together on one problem ([22], [7]),
- *Cultural and language differences*, often having a negative impact on communication and coordination ([7], [28], [14]), and
- *Lack of trust* due to difficulties in building personal relationships between sites or perceived rivalries ([8], [42]).

Problems at one site mainly occur when the personnel on the site does not match the requirements of the tasks assigned to the site, i.e., if people do not have the right qualifications. This could lead to poor performance at this site. Typical reasons are:

- *Skill and experience* that can often not be judged accurately at remote sites ([23]) or
- *Insufficient knowledge* at development sites, e.g., if domain knowledge has to be built up at

sites ([27], [23]) or if there are cultural differences (e.g., Indian developers writing a user interface for a European application) ([28]).

## 2.3. Tactics and Strategies for Distributing Work

Several tactics are applied in practice in order to avoid these problems, both during initiation and management of GSD projects. In terms of assignment of tasks to sites, different tactics are suggested:

In order to avoid problems between sites, the main tactic is minimization of collaboration needed between sites, since this minimizes the negative impact of communication and coordination problems. This can be achieved by minimizing coupling, i.e., the dependencies between tasks assigned to different sites. ([7], [33], [29]).

Other tactics for reducing problems between sites aim at minimizing the differences between sites, e.g., by reducing the time shift ([7]) or cultural differences ([28]).

Problems at sites can be reduced by matching the requirements of the tasks with the knowledge and experience of the sites they are assigned to, e.g., by assigning tasks that require certain domain knowledge to sites possessing this knowledge ([23]) or by only assigning "culturally neutral" tasks to sites with different cultures ([28]).

However, most of the tactics and strategies reported have a very strong focus on the management of GSD during project execution and do not consider the task assignment during project initiation ([29], [28], [7], [36], [23], [34]).

Other strategies for distributing work come from the area of organizational theory: The contingency theory suggests different strategies for assigning work and defining communication channels and also considers globally distributed organizations [6], [17]. However, this is usually done on a relatively high level and does not consider the specifics of GSD. Grinter et al. [20] applied this to distributed development and identified four methods of organizing work in GSD: functional areas of expertise, product structure, process steps, and customization. Each method comes with a strategy for task assignment and with different benefits and costs.

## 3. Requirements for a Distribution Model

A distribution model that properly models GSD should at least address all relevant concepts introduced in section 2: a set of goals, resources available at sites (or nodes), required work divided into tasks, and a mapping that describes the potential impact of every assignment to the goals. However, the special problems and characteristics of GSD require a set of additional properties for a distribution model:

**REQ1:** Multi-objective goal function: As shown in section 2.2, there exists a variety of different goal types for distributed development projects, not just aiming at cost minimization. A distribution model should be able to evaluate a distribution according to a set of (potentially conflicting) goals.

**REQ2:** Properties of tasks and sites: The model must be able to describe the specific characteristics of tasks and sites. These characteristics can reflect the problems at sites as described above (e.g., certain knowledge is required by a task and not available at a site). For tasks, these are mainly the requirements that need to be fulfilled by sites to perform the tasks. For sites, these are the assumed context factors relevant for assessing the ability to perform tasks at sites. It should also be possible to define constraints, e.g., if a certain task must not be done at a certain sites due to political reasons.

**REQ3:** Dependencies between tasks and between sites: It must be possible to characterize and model relationships between tasks and between sites (e.g., high communication intensity between two tasks, low communication efficiency between two sites).

**REQ4:** Adaptability: Since for every development project and environment, specific properties are of importance and specific goals are to be met, it must be possible to adapt and change the goals, properties, and dependencies within the model and to introduce new goal functions.

**REQ5:** Formality: The model must have an appropriate degree of formality in order to be able to apply formal algorithms.

**REQ6:** Empirically based criteria: The mappings (i.e., the functions that describe the effects of tasks and site characteristics on relevant process or product properties) need to be defined as empirical relationships. Due to the nature of software development, many tasks are human-based and non-deterministic so that it is necessary to integrate empirical relationships (as real evidence or as hypothesis) into the distribution model.

## 4. Existing Task Distribution Approaches

In this section, selected existing distribution models are presented that assign tasks to sites in a network. For every approach, its feasibility in using it as a distribution model for a decision support system for GSD is analyzed by evaluating the requirements of section 3.

Three different domains were selected that contain possibly useful decision models:

*Distributed Software Development:* Some approaches for modeling or optimizing aspects of distributing software development tasks already exist. They have the advantage of already being in the targeted domain and therefore it is possible to model specific aspects of GSD.

*Distributed production:* Working together on one project at different sites or companies has a long tradition in the production domain. Therefore, different optimization approaches for distributing production tasks exist. Many of these approaches, such as operations research approaches, are very formal and come with algorithms for finding optimal distributions.

*Distributed systems:* Even if this domain handles tasks within and between computers and not humans, it should be regarded, too: A lot of research was spent on models that find optimal assignments of computing tasks (e.g. threads, files) to nodes (e.g. processors, computers) that can be regarded as sites. Many of these models are multi-objective and formal, and already offer algorithms for finding optimal solutions.

A fourth domain that could also be considered is the area of distributed research: Like software development, R&D activities can be done in a distributed and global way. As shown in [19], the effective selection of partners and the distribution of activities are of importance here, too. However, since there is no established research on distribution models for this domain, it will not be regarded here.

In the following section, typical approaches from these domains are discussed in more detail. After that, they are compared based on the previously stated requirements.

### 4.1. Distributing Software Development Tasks

Mockus and Weiss ([33]) developed a model for optimizing work assignments that deals with Modification Requests, i.e., a set of changes to existing files. The model is used as basis for an optimization algorithm that identifies sets of files that should be transferred to remote sites in order to minimize the work pieces spanning multiple sites. The main idea here is that these increase the need for cross-site communication and thereby decrease the overall productivity. Therefore, minimizing this multi-site work maximizes productivity. Based on the empirical data for every modification request, the algorithm knows the multi-site requests. It iteratively selects sets of files, transfers them to other sites, and calculates the multi-site effort again in order to find better distributions.

The underlying model regards only one goal, the minimization of multi-site modification requests with free resources as constraints; it is therefore not multi-objective. Both tasks (i.e. the responsibility over a file) and sites have properties; however, they only reflect the needed, respectively the available, resources and do not consider knowledge and experience. Dependencies exist only between sites (number of multi-site modification requests). The model is not completely formal, but formal enough to provide an optimization algorithm. Adaptability is not considered. Only one empirically based criterion for optimization is used (number of multi-site requests). In earlier publications, a decrease in productivity was shown by empirical means.

In the Global Studio Project ([34], [38]), a more practical distribution model was created for the optimization of work assignments. In this project, teams from different universities worldwide practiced methods and techniques of distributed development. The total development work was split up into coherent packages of work together with their required knowledge and temporal dependencies (i.e., which packages had to be done before other packages). Based on this data and the knowledge of the teams, the packages were assigned manually.

No goal function was modeled explicitly. The underlying goals were an optimal distribution of the workload among the teams and the building of specialized knowledge in the different teams. Both work packages and teams had properties reflecting the needed (respectively available) resources and knowledge. Temporal dependencies between tasks were modeled; between teams, no dependencies existed. The model has no formality, since it was only used manually and for communication purposes and is not explicit. Adapting the model is therefore hardly possible. The criteria for task allocation come from the experiences of a previous run of the project and were evaluated in the second run. They are therefore empirically based, but not on a broad basis.

Madachy [32] developed a cost model for estimating effort in distributed projects. These estimations can also be used for estimating total costs, the distribution of effort among sites, and the personnel needs at each site. The model is an extension of the COCOMOII approach [4] used for estimating effort. COCOMO is based on historical data, the estimated product size, and a set of influence factors. In this model, the total effort is divided into different phases; the phases are distributed among the sites; and for each phase at each site the effort is adjusted according to the site characteristics.

Estimated effort or estimated costs can be used as a goal function that is not multi-objective. Sites have various properties (e.g. capability, experience of different roles). Tasks also have different properties, e.g., the required roles within a phase at a certain site. However, no dependencies between sites or tasks are modeled. Adding new properties of nodes is intended in the model and very easy. Adding new objectives or adding dependencies is not possible. The model is not completely formal, but formal enough to provide an optimization algorithm. Some of the criteria come from COCOMO, which is empirically based to a very large extent. However, other criteria and factors are not explicitly named; the model is rather a framework for inserting one's own criteria. It therefore does not come with many criteria for task allocation.

Setamanit, Wakeland, and Raffo ([41], [44]) developed a model for the simulation of GSD. It is intended for studying different alternatives for GSD and for supporting task allocation and site selection decisions. The model is hybrid with discrete-event and system dynamics sub-models. It contains several site-specific sub-models that reflect the special resources and capabilities at all sites. These site-specific models are enhanced by an interaction effect model that adjusts the results depending on the communication efficiency between sites.

The model considers different goals such as effort and defect rate. There are different properties of sites such as available resources and individual productivity, but no properties of tasks. Dependencies between sites are also contained in the model, e.g., cultural difference or distance. However, there are nearly no dependencies between tasks. It has formality since it is built on simulation paradigms. Adaptability was not intended but is probably possible. The criteria and factors of the model were based on a literature review but not on specific empirical studies. The model was validated in individual projects but not on a broad level.

Prikladnicki et al. [37] defined a reference model for global software development that also contains a process for task allocation. It defines a set of different criteria for evaluating available development sites for a specific project. This includes a detailed risk and cost-benefit analysis for every site in order to identify the best sites for development.

The reference model supports multiple goals and is also adaptable. Properties of tasks and sites are regarded by the criteria defined for site selection. Dependencies between sites are considered as well; however, dependencies between tasks do not appear in the criteria. The model lacks formality, thus making it difficult to apply optimization algorithms. The model seems to be empirically based and was validated in case studies, but it is not clear whether the proposed criteria stem from empirical studies.

There are other approaches for modeling GSD ([15], [39]), but as they focus only on some selected aspects, they cannot be a basis for decision support systems.

## 4.2. Distributing Production Tasks

The problem of production distribution has been in the focus of research for decades and has led to various distribution models ([46]). Usually, the main problem here is finding an optimal assignment of production work to plants, whereas optimality means in most cases cost minimization.

A distribution model for steel production is presented by Chen and Wang [9]. The model consists of a set of nodes that represent both production sites and regional sales markets. The goal is to maximize the profit that is calculated by the sales in all regions and the transport and production costs. Linear programming is used for this model: A goal function representing the profit is to be maximized while the capacities in the factories and the sales forecast formulate the constraints. The cost depends on the distribution of the production across the sites.

As the model only regards costs, it is not multi-objective. Production sites do not have properties except their free resources. Tasks possess properties; they are described by their amount and the raw material needed. Dependencies exist only indirectly between sites as transportation costs between a site and the central plant. Between tasks, there are no dependencies. The model cannot be extended because adding new dependencies or goals is not possible in a linear programming model. However, due to its origin in linear programming, it is completely formalized and optimal assignments can be found easily.

A similar approach is presented by Cohen and Moon ([11]). It consists of a set of plants, where the production takes place, and distribution centers, where the products are delivered to. The goal is minimization of overall costs while satisfying the demand at the distribution centers. However, this model is defined by more variables and can be optimized for more criteria such as the product flow from each plant to each distribution center and the production distribution across several production stages. Solving the model is thus harder and requires more complex algorithms.

The model again only considers costs and therefore is not multi-objective. Both tasks and sites have properties that are expressed in matrices, naming for instance the production costs for each product at each plant. Dependencies do not exist between tasks and only between plants and distribution centers, but not directly between sites. The model is not intended for being adapted but is again completely formalized.

A model used as the basis for a decision support system for the worldwide distribution of the car production at BMW is presented in [16]. Here the goal is also minimization of the overall production costs, but investment planning is regarded as well. The model considers the use and supply of material and engines needed at the sites. As for the other models, a cost function depending on the assignment of productions to sites is to be minimized. Demands and supplies of products and material in different regions are formalized as restrictions for the function.

The model is not multi-objective, since it again only considers costs. However, since many aspects such as use of material and investment planning are modeled as costs, it is slightly multi-objective. Both tasks (different products with different demand on material) and sites have properties. Dependencies between sites exist only indirectly as transportation costs from regions to sites. Dependencies between tasks are not part of the model. Adding new properties might be possible. However, adding dependencies or new goals cannot be supported by the model. It is again formalized.

### 4.3. Distributing Computing Tasks

In a distributed system, different computers or processors run one application. One central question in the design of distributed applications is the decision of which tasks to do where. Many distribution models have been developed in order to solve different assignment problems.

The problem of processor allocation, i.e., the optimal assignment of computation tasks to processors in a network, is very common in distributed operating systems. It is often solved by using graph algorithms ([2]). Bokhari developed an algorithm that assigns software modules (i.e., tasks) to nodes with the objective of minimizing the weighted sum of communication and execution costs ([5]). The modules are assumed to communicate in a tree-like structure, a so called "invocation tree". Based on this assumption, the model creates an assignment tree showing the assignment of modules to processors. It goes along with an algorithm that finds an optimal allocation with respect to minimal communication and execution time. A more detailed description of the model can be found in the appendix.

Tasks can be assigned to all nodes, but since a tree structure is required between tasks, there is no complete freedom in the structure of tasks. The algorithm minimizes the weighted sum of communication and execution costs. It is therefore multi-objective. Properties of tasks and nodes are the needed, respectively the provided, computing power, so both are part of the model. Dependencies exist as communication needs between modules and as speed between nodes. However, the tree structure between nodes restricts these dependencies. The model was not intended for being adapted, but adding new properties, dependencies, and goals may be possible. It is also much formalized.

Dynamic assignment of tasks to computers in a cluster is another well-known problem in cluster or grid computing. For dynamic job assignment, Amir et al. developed a distribution model ([1]). Based on the expected resource usage of incoming jobs, it assigns the tasks with the objective of minimizing the overall slowdown. The main idea of Amir et al. is the aggregation of the different resource needs (e.g., CPU, memory) of a job into "opportunity costs" for each computer. The job is then assigned to the computer with the least costs.

As different objectives are aggregated into costs, the model is multi-objective. Both tasks and nodes have properties (CPU power, main memory) that influence the opportunity costs. However, dependencies are not modeled. Adapting the model by extending the goal function or the properties of tasks and nodes is easy by just extending the opportunity cost function. Adding new dependencies is not possible. Again, the model is formalized.

The "File Allocation Problem" (FAP) deals with the assignment of files to nodes in a computer network. The goal is to find an optimal allocation that minimizes the needed storage costs while maximizing the update and access speed for files. Chu developed a distribution model for this problem ([10]). Here cost-optimal distributions of plants and warehouses are discussed, similar to the approaches presented in section 4.2. Chu presents a mathematical distribution model for the FAP that aims at minimizing the operating costs. These costs are an aggregation of storage and transmission costs, both weighted by prices in $. The model is therefore multi-objective.

Both computers and files have properties influencing the goals, e.g., frequency of modification of a file or request rate for a file at a computer. Dependencies exist between nodes (communication speed between computers), but no dependencies between tasks are modeled. The model was not intended for being extended. Adding new properties and their influences on goals may be possible, but adding new objectives or dependencies would destroy the model. It is mathematically formalized.

## 5. Comparison of the Approaches

The comparison of the different modeling approaches is shown in Table 1.

The modeling approaches already existing within the domain of distributed software development usually focus only on small aspects of the distribution. Thus, none of them regards both properties of and dependencies between both the nodes in the network and the tasks. Most of them are not truly multi-objective, ei-ther; only isolated goals such as effort or the number of multi-site work items are considered.

**Table 1. Comparison of the different distribution models and their fulfillment of the requirements: Not (-), partly (o), mostly (+) or totally (++) fulfilled (gray: not comparable)**

| Domain | Approach | Multi-objective goal | Properties Tasks | Properties Nodes | Dependencies Tasks | Dependencies Nodes | Adaptability | Formality | Empirically based |
|---|---|---|---|---|---|---|---|---|---|
| Distributed SW Development | Modification Requests [33] | - | o | o | - | + | - | + | o |
| Distributed SW Development | Global Studio Project [34], [38] |  | + | + | + | - | - | - | + |
| Distributed SW Development | Distributed CoCoMo [32] | - | + | ++ | - | - | o | + | o |
| Distributed SW Development | Simulation Model [44], [41] | + | - | + | o | + | o | + | o |
| Distributed SW Development | Reference Model for GSD [37] | + | + | + | - | + | + | - | o |
| Distributed Production | Linear Programming [9] | - | + | o | - | o | - | ++ |  |
| Distributed Production | Plant Distribution [11] | - | + | + | - | - | - | ++ |  |
| Distributed Production | BMW Production [16] | - | + | + | - | o | - | ++ |  |
| Distributed Systems | Processor Allocation [5] | o | + | + | o | + | o | + |  |
| Distributed Systems | Opportunity Costs [1] | ++ | + | + | - | - | o | + |  |
| Distributed Systems | File Allocation [10] | + | + | + | - | + | - | ++ |  |

Distribution models within the production-distribution domain are, in general, more formalized than in distributed software development, since they are based on operations research methods. However, these models only focus on cost minimization and are therefore not multi-objective. Besides, they have a strong focus on modeling production facilities while neglecting the details of tasks. Both the distribution models in software development and distributed production are usually made for specific environments. They therefore lack adaptability.

A general problem of the models in production and distributed systems is their obvious lack of empirical evaluation with respect to GSD specifics. But leaving that fact aside, the modeling approaches of the distributed systems domain seem to fit best to the requirements: All regarded models are multi-objective, since they try to optimize both the costs at the nodes and the communication between them. The distribution model of processor allocation is the only one that considers both properties of and dependencies between both tasks and nodes, even if it is restricted to certain dependencies between the tasks. Since most of the distribution models in distributed systems result in an optimization algorithm executable by machines, they are all quite formalized. However, these distribution models usually are also difficult to adapt.

## 6. Conclusion and Future Work

As the comparison of different existing distribution models within the domains of distributed software development, distributed production, and distributed systems shows, there exists no single model that fulfills all stated requirements. However, the analysis of the models and their underlying principles reveals several ideas that could be transferred to the problem of distributing GSD tasks.

The models developed in the distributed systems domain match the requirements to a large extent. But they do not consider specific empirically proven facts and criteria of GSD. It is also hard to extend them, as they were usually not intended for that purpose.

However, a very promising model seems to be the processor allocation in Bokhari's graph algorithm, as it is the only model that fulfills every requirement at least partly. In future work, we plan to develop a model that can be mapped to the processor allocation. In the appendix, we roughly sketch an initial model. This model should use the empirically based criteria of the already existing models for GSD or further empirical evidence that will be generated in the future.

Another aspect that was shown by comparing the different distribution models is the fact that all models focus on some special aspects: There is no model that considers the properties of and dependencies between tasks and nodes, is multi-objective and adaptable at the same time. This fact suggests that a model covering all these requirements probably might be too complex and not realizable.

One question that arises when it comes to the development of a model for GSD is: Could it be possible to create a quantitative model for predicting and optimizing human activities? Humans act and work in a far more complex way than, for example, distributed com-puter networks, and it is never possible to consider all influence factors when modeling their work. This has to be investigated in future research; and it will probably not be possible to develop a model as accurate as the ones in distributed production or distributed systems. However, there exist different other approaches that model human behavior successfully in a very formal way ([35], [30]) and we believe that even approximate results will help to improve the task assignment decision

Future research has to handle a lot of other questions that could not be handled in this comparison. Some of these are: How can organization-specific distribution scenarios be described in order to derive distribution requirements and criteria systematically? How to handle time? Planning software projects includes much time planning, as some resources are only available at some time or as the project time is one of the goals. What are further requirements for a DSS? The previously stated requirements are mainly derived from the literature and should be elaborated further.

## Appendix: Using Processor Allocation for GSD

As presented before, the model used in Bokhari's algorithm for assigning tasks to processors [5] seems to fit to a large extent to the requirements for a GSD distribution model. In the following, we will sketch a possible adoption for GSD.

The main element of the model is the *invocation tree* of the tasks to be assigned and the dependencies between them. Its tree structure stems from the fact that a module usually invokes a set of other modules it communicates with (and that do not communicate among each other). The model gets a set of variables and functions as inputs:

$e_{ip}$: cost of executing module i on processor p

$d_{ij}$: amount of data transmitted between module i and module j (i.e., the required communication).

$s_{pq}$: cost of transmitting one unit of data between processors p and q.

$S_{pq}(d_{ij})$: cost of transmitting data between modules i and j if i is assigned to processor p and j is assigned to q.

The minimal cost assignment is found by building an *assignment graph* from the invocation tree using the following rules:

- For each node i in the invocation tree (i.e., a module) and every processor p, the assignment graph contains a node (i, p). The graph thus contains a layer (i, 1), (i, 2)… of nodes for every node in the tree.
- If two nodes in the tree have an edge between them, the graph contains edges between all nodes of the corresponding layer.
- Edges between nodes (i, p) and (j, q) have weights $e_{jq} + S_{pq}(d_{ij})$.
- A source node is connected to all nodes (1, 1), (1, 2)… of the first layer with edges of weight 0.
- All nodes in the graph that correspond to leaves in the tree are connected to terminal nodes with edges of weight 0.

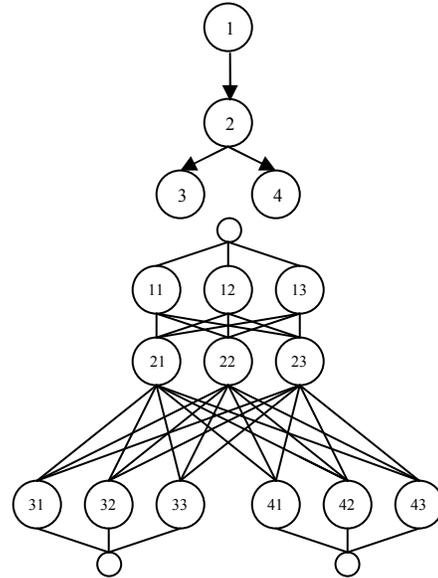

**Figure 1.** An invocation tree and the resulting assignment graph for three processors.

The problem of finding an optimal assignment is then reduced to finding a minimal (with respect to the sum of the edge weights) tree that connects every terminal node with the source node. This can be done using dynamic programming in $O(mn^2)$, with m being the number of modules and n the number of processors.

If this model is adapted to GSD, the biggest problem is the constraint of the tasks being restricted to a tree-like structure. This forbids modeling arbitrary dependencies between tasks. However, software processes following a waterfall model can be modeled as a chain of sequential tasks.

Then the input parameters can be regarded as:

$e_{ip}$: cost of assigning task i to site p (depending on the properties of i and p)

$d_{ij}$: needed communication between task i and task j

$s_{pq}$: dependencies between site p and site q (e.g., communication efficiency)

$S_{pq}(d_{ij}, s_{pq})$: communication cost between tasks i and j if i is assigned to site p and j is assigned to q.

These functions and parameters have to be further specified, which is not supported by the model as it does not come from the GSD domain. However, the experiences and characteristics of the specific GSD models presented can be used here.

Another problem of this adoption is the fact that Bokhari's algorithm is not really multi-objective. It is therefore only possible to optimize one single goal. This could be solved by identifying optimal solutions for every single goal and then comparing them. Another way might be the use of the opportunity cost approach in [1] or the use of goal programming from operations research.

## References

[1] Amir, Y., Awerbuch, B., Barak, A., Borgstrom, R.S., Keren, A.: "An Opportunity Cost Approach for Job Assignment in a Scalable Computing Cluster". *IEEE Transactions on parallel and distributed Systems*, Vol. 11, no. 7, July 2000

[2] Barbosa, V. C.: *An Introduction to distributed Algorithms*. MIT Press, 1996

[3] Bass, M., Paulish, D.: "Global Software Development Process Research at Siemens". *Third International Workshop on Global Software Development* ICSE 2004

[4] Boehm, B., Abts, C., Brown, A., Chulani, S., Clark, B., Horowitz, E., Madachy, R., Reifer, D., Steece, B.: *Software Cost Estimation with COCOMO II*. Prentice-Hall (2000)

[5] Bokhari, S. H.: "A Shortest Tree Algorithm for Optimal Assignments across Space and Time in a distributed Processor System". IEEE *Transactions on Software Engineering* SE-7:6, 583-589 (1981)

[6] Burton, R. M., Obel, B.: *Strategic Organizational Diagnosis and Design: Developing Theory for Application*, 2nd ed. Norwell, MA: Kluwer, 1998.

[7] Carmel, E., Agarwal, R.: "Tactical Approaches for Alleviating Distance in Global Software Development". IEEE *Software* Volume 18, Issue 2 (March 2001)

[8] Casey, V., Richardson, I.: "Uncovering the Reality within Virtual Software Teams". *International workshop on Global software development for the practitioner*, 2006

[9] Cheng, M., Wang, W.: A linear programming model for integrated steel production and distribution planning. *International Journal of Operations & Production Management*, Vol. 17 No. 6, 1997

[10] Chu, W. W.: "Optimal File Allocation in a Multiple Computer System". IEEE *Transactions on Computers* C-18:10, 885-889 (1969)

[11] Cohen, M.A., Moon, S.: "An integrated plant loading model with economies of scale and scope". *European Journal of Operational Research* Vol. 50 Issue 3 (1991)

[12] Damian, D.: "Stakeholders in Global Requirements Engineering: Lessons Learned from Practice". IEEE *Software* March/April 2007 (Vol. 24, No. 2)

[13] Damian, D., Moitra, D.: "Global Software Development: How Far Have We Come?" IEEE *Software*, 23(5):17-19. (2006)

[14] Dubie, D.: "Outsourcing Moves Closer to Home". *CIO Today* December 18, 2007

[15] Espinosa, J. A., Carmel, E.: "Modeling Coordination Costs Due to Time separation in Global Software Teams". *International Workshop on Global Software Development*, 2003.

[16] Fleischmann, B., Ferber, S., Henrich, P.: "Strategic Planning of BMW's Global Production Network". *Interfaces* Vol. 36, No. 3, May–June 2006, pp. 194–208

[17] Galbraith, J.R.: *Designing the Global Corporation*, San Francisco: Jossey-Bass, 2000

[18] Gareiss, R.: "Analyzing the Outsourcers". *Information Week* November 18, 2002

[19] Gerybadzea, A., Reger, G.: Globalization of R&D: recent changes in the management of innovation in transnational corporations. *Research Policy*, 18(2-3): 251-274 (1999)

[20] Grinter, R.E., Herbsleb, J.D., Perry, D.E.: The geography of coordination: Dealing with distance in R&D work. *Proc. ACM Conference on Supporting Group Work* 1999 (GROUP 99)

[21] Herbsleb, J.D.: "Global Software Engineering: The Future of Socio-technical Coordination". *Proceeding of the 29th international conference on Software engineering* ICSE 2007

[22] Herbsleb J. D., Grinter, R. E.: "Splitting the organization and integrating the code: Conway's law revisited". *Proceedings of the 21st International Conference on Software Engineering* ICSE '99

[23] Herbsleb, J., Paulish, D.J., Bass, M.: "Global software development at Siemens: Experience from nine projects". *Proceedings of the 27th International Conference on Software Engineering* ICSE 2005, St. Louis, MO, May 15-21, pp. 524-533.

[24] Herbsleb, J.D., Mockus, A., Finholt, T.A., Grinter, R.E.: "An empirical study of global software development: Distance and speed". *Proceedings, International Conference on Software Engineering*, Toronto, Canada 2001

[25] Herbsleb, J. D., Moitra, D.: "Guest editors' introduction: Global software development". IEEE *Software* Volume 18, Issue 2 pp.16-20. (2001)

[26] Ju, D.: "A Concerted Effort towards Flourishing Global Software Development". *Proc. International workshop on Global software development for the practitioner*, 2006

[27] Kobitzsch, W., Rombach, H.D., Feldmann, R.L.: "Outsourcing in India". IEEE *Software*, Volume 18, Issue 2 (2001), pp. 78-86

[28] Krishna, S., Sahay, S., Walsham, G.: "Managing cross-cultural issues in Global Software Outsourcing". *Communications of the ACM* Volume 47, Issue 4 (Apr. 2004).

[29] Lee, G., DeLone, W., Espinosa, J. A.: "Ambidextrous coping strategies in globally distributed software development projects". *Communications of the ACM*, Volume 49, Issue 10 (October 2006)

[30] Levchuk, G. M., Levchuk, Y., N., Luo, J., Pattipati, K., R., Kleinman, D. L.: Normative Design of Organizations—Part II: Organizational Structure. *IEEE Tansactions on Systems, Man, and Cybernetics — Part A: Systems and Humans*, vol. 32, no. 3, May 2002

[31] Lindqvist, E., Lundell, B., Lings, B.: "Distributed Development in an Intra-national, Intra-organisational Context: An Experience Report". *Proc. International workshop on Global software development for the practitioner*, 2006

[32] Madachy, R.: "Distributed Global Development Parametric Cost Modeling". *Proceedings International Conference on Software Process*, ICSP 2007, Springer-Verlag (2007)

[33] A. Mockus, D. M. Weiss: "Globalization by Chunking: A Quantitative Approach". IEEE *Software* Volume 18, Issue 2 (March 2001)

[34] Mullick, N., Bass, M., Houda, Z., Paulish, D.J., Cataldo M., Herbsleb, J. D., Bass, L.: "Siemens Global Studio Project: Experiences Adopting an Integrated GSD Infrastructure". *International Conference on Global Software Engineering* ICGSE 2006

[35] Pete, A.: *Structural Congruence of Tasks and Organizations*. PhD Thesis, University of Connecticut, 1995

[36] Pilatti, L., Audy, J., Prikladnicki, R.: "Software Configuration Management over a Global Software Development Environment: Lessons Learned from a Case Study". *Proc. International workshop on Global software development for the practitioner*, 2006

[37] Prikladnicki, R., Audy, J. L. N., Evaristo, R.: A Reference Model for Global Software Development: Findings from a Case Study. ". *International Conference on Global Software Engineering* ICGSE 2006

[38] Raghvinder, S., Bass, M., Mullick, N., Paulish, D. J., Kazmeier, J.: *Global Software Development Handbook*. Auerbach 2006

[39] Ramasubbu, N., Balan, R. K.: "Globally Distributed Software Development Project Performance: An Empirical Analysis". *Proc. 6th joint meeting of the European software engineering conference and the ACM SIGSOFT symposium on the foundations of software engineering* (2007)

[40] Scott-Morton, M. S.: *Management Decision Support Systems: Computer-Based Support for Decision Making*. Cambridge, MA, 1971

[41] Setamanit, S., Raffo, D.: Identifying Key Success Factors for Globally Distributed Software Development Project Using Simulation: A Case Study. *International Conference on Software Process*, ICSP 2008

[42] Smite, D., Moe, N. B.: Understanding a Lack of Trust in Global Software Teams: A Multiple-Case Study. *PROFES 2007*: 20-34

[43] Seshagiri, G.: "GSD: Not a Business Necessity, but a March of Folly". IEEE *Software*, 23(5):63f. (2006)

[44] Setamanit, S., Wakeland, W.W., Raffo, D.: Using simulation to evaluate global software development task allocation strategies. *Software Process: Improvement and Practice* 12(5): 491-503 (2007)

[45] Turban, E.: *Decision Support and Expert Systems*. 4th ed., Prentice Hall, 1995

[46] Vidal, C., Goettschalkx, M.: Strategic production-distribution models: A critical review with emphasis on global supply chain models. *European Journal of Operational Research*. Vol. 98 Issue 1 (1997)